\documentstyle[aps,epsfig]{revtex}
\begin{document}
\twocolumn[\hsize\textwidth\columnwidth\hsize\csname
@twocolumnfalse\endcsname

\title{$SL(2,R)$ model with two Hamiltonian constraints}

\author{Merced Montesinos$^{a,b,c,1}$, 
Carlo Rovelli$^{a,c,2}$
and Thomas Thiemann$^{d, 3}$
}
\address{\vskip.1cm
$^a$ Centre de Physique Th\'eorique, CNRS Luminy, F-13288 Marseille, 
France.\\
$^b$ Departamento de F\'{\i}sica, Centro de Investigaci\'on y de 
Estudios Avanzados del I.P.N.,\\ 
Av. I.P.N. No. 2508, 07000 Ciudad de M\'exico, M\'exico.\\
$^c$ Department of Physics and Astronomy, 
University of Pittsburgh, Pittsburgh, PA 15260, USA.\\
$^d$ Albert-Einstein-Institut, MPI f.\ Gravitationsphysik, 
Schlaatzweg 1, 14473 Postdam, Germany.\\
\vskip3pt{ $^1${merced@fis.cinvestav.mx},\  
$^2${rovelli@cpt.univ-mrs.fr},\ $^3${thiemann@aei-postdam.mpg.de}} 
}  

\maketitle \vskip10pt

\begin{abstract} 

We describe a simple dynamical model characterized by the 
presence of \textit{two} noncommuting Hamiltonian constraints.  
This feature mimics the constraint structure of general 
relativity, where there is one Hamiltonian constraint associated 
with each space point.  We solve the classical and quantum dynamics 
of the model, which turns out to be governed by an $SL(2,R)$ 
gauge symmetry, local in time.  In classical theory, we solve 
the equations of motion, find a $SO(2,2)$ algebra of Dirac 
observables, find the gauge transformations for the Lagrangian 
and canonical variables and for the Lagrange multipliers.  In 
quantum theory, we find the physical states, the quantum 
observables, and the physical inner product, which is determined 
by the reality conditions.  In addition, we construct the 
classical and quantum evolving constants of the system.  The 
model illustrates how to describe physical gauge-invariant 
relative evolution when coordinate time evolution is a gauge.

\end{abstract}
\pacs{PACS: 04.60.Ds.}
\vskip 5pt]


\section{Introduction}

General relativity (GR) has a characteristic gauge invariance, 
which implies that its canonical Hamiltonian vanishes weakly.  As 
a consequence, its dynamics is not governed by a genuine 
Hamiltonian, but rather by a ``Hamiltonian constraint''.  This 
peculiar feature of the theory has a crucial physical 
significance, connected to the relational nature of the 
general-relativistic spatiotemporal notions 
\cite{diff,Rovelli99,models}, and raises a number of important 
conceptual as well as technical problems, particularly in 
relation to the quantization of the theory \cite{quantization}.  
In the past, much clarity has been shed on these problems by 
studying finite dimensional models mimicking the constraint 
structure of the theory, and in particular, having weakly 
vanishing Hamiltonian \cite{models}.

There is an aspect of the constraint structure of GR, however, 
which, as far as we are aware, has not been analyzed with the 
use of constrained models.  In GR, there isn't just a single 
Hamiltonian constraint, but rather a \textit{family\/} of 
Hamiltonian constraints, one, so to say, for each 
coordinate-space point.  Furthermore, the Hamiltonian constraints 
do not commute with each other (have nonvanishing Poisson 
brackets with each other).  Indeed, the constraint algebra of GR 
has the well known structure
\begin{equation}
		\{H, H\} \sim  D, \ \ \ 
		\{H, D\} \sim  H, \ \ \ 
		\{D, D\} \sim  D, 
	\label{dd}
\end{equation} 
where $H$ represents the Hamiltonian constraints and $D$ the 
diffeomorphism constraints.  In this paper we present a model 
that mimics this aspect of GR.

The model we present has three constraints, which we call 
$H_{1}$,  $H_{2}$ and $D$. Their algebra has the 
structure 
\begin{equation}
		\{H_{1}, H_{2}\} \sim  D, \ \ \ 
		\{H_{i}, D\} \sim  H_{i}, 
	\label{hd2}  
\end{equation} 
which mimics (\ref{dd}). (Models with several {\it 
commuting\/} Hamiltonian constraints were considered in 
\cite{luca}.) The constraints $H_{1}$ and $H_{2}$ are quadratic 
in the momenta, while $D$ is linear, as their correspondents in 
GR. 

The model has an interesting structure which exemplifies in a non 
trivial manner various aspects of the quantization and 
interpretation of the fully constrained systems.  We analyze in 
detail its classical and quantum dynamics, which can both be 
solved completely.  We display the general solution of the 
equations of motion and the finite gauge transformation of 
variables and Lagrange multipliers.  The constraint algebra turns 
out to be $SL(2,R)$ and the model is invariant under an $SL(2,R)$ 
gauge invariance, local in time.  We find a complete $SO(2,2)$ 
algebra of gauge invariant observables, as well as a (smaller) 
complete set of independent observables.  The phase space turns 
out to have the topology of four cones connected at their 
vertices.  We then study the quantum dynamics, solve the Dirac 
constraints, exhibit the physical states explicitly, and 
construct a complete family of gauge invariant operators.  The 
reality properties of the gauge invariant operators fix uniquely 
the physical scalar product.  In addition, we define the 
classical and quantum evolving constants \cite{evolving} of the 
system, and we discuss the observability of evolution for the 
systems (like GR) in which time is a gauge and the theory has no 
preferred physical time.


\section{Classical Dynamics}

{\it Definition of the model.}
The model we consider is defined by the action 
\begin{eqnarray}
& S[{\vec u},{\vec v}, N, M , \lambda ] = \hspace{14em}  &
\nonumber \\
& \frac12 
{\displaystyle \int}\ dt \left[\, N\, ({\cal D} \vec u^2 + \vec v^2)
+ M\, ({\cal D} \vec v^2 + \vec u^2)\, 
\right], &\label{Action} 
\end{eqnarray}
where
\begin{equation}
{\cal D} {\vec u}  =  \frac{1}{N} (\dot {\vec u} - 
\lambda {\vec u}), \ \ \ \ \ 
{\cal D} {\vec v}  =  \frac{1}{M} (\dot {\vec v} + 
\lambda {\vec v}); 
\end{equation}
the two Lagrangian dynamical variables ${\vec u}=(u^1,u^2)$ and 
${\vec v}=(v^1,v^2)$ are two-dimensional real vectors; $N$, $M$ 
and $\lambda$ are Lagrange multipliers.  The squares are taken in 
$R^{2}$: $\vec u^2 = \vec u\cdot \vec u = 
(u^{1})^{2}+(u^{2})^{2}$.

{\it Hamiltonian analysis.} 
The Hamiltonian analysis is simplified by first rewriting 
the action in the following form
\begin{eqnarray}
S  &=&  \int dt \left [ \dot {\vec u} \cdot  {\cal D} {\vec u} + 
\dot {\vec v} \cdot {\cal D} {\vec v} - 
 N \frac12 ({\cal D}{\vec u}^2 -\vec v^2) \right.  \nonumber \\
&& \left. \hspace{2em} -  M \frac12 ({\cal D}{\vec v}^2 -\vec u^2) 
- \lambda ({\vec u}\cdot {\cal D}{\vec u} 
- {\vec v} \cdot {\cal D}{\vec v}) \right ] \, .\label{Monica}
\end{eqnarray}
From this form, we see that the momenta conjugate to 
$\vec u$ and $\vec v$ are  
\begin{equation}
{\vec p}  =  {\cal D}{\vec u} \hspace{2em} \mathrm{and} 
\hspace{2em}
{\vec \pi} =  {\cal D}{\vec v} 
\end{equation}
respectively, and that we have a weakly vanishing Hamiltonian and  
three primary constraints
\begin{eqnarray}
H_1 & = & \frac12 (\vec p^2 -\vec v^2)\, ,\nonumber\\
H_2 & = & \frac12 (\vec{\pi}^2 -\vec u^2)\, ,\nonumber\\
D\ & = & {\vec u}\cdot {\vec p} - 
{\vec v} \cdot {\vec \pi}\, . \label{Hamiltonian} 
\end{eqnarray}
The Hamilton equations of motion are 
 \begin{eqnarray}
\dot {\vec u} =  N {\vec p} +\lambda {\vec u}\, , 
&\hspace{3em}&
\dot {\vec v}  =  M {\vec \pi} - \lambda {\vec v} \, ,
\nonumber\\
\dot {\vec p} =  M {\vec u} -\lambda {\vec p}\, , 
&\hspace{3em}&
\dot {\vec \pi}  =  N {\vec v} +\lambda {\vec \pi} \, . 
\label{Equations}
\end{eqnarray}

Using (\ref{Hamiltonian}) and (\ref{Equations}) we find the evolution of 
the constraints 
\begin{eqnarray}
{\dot H}_1 & = & \ M D -2\lambda H_1\, ,\nonumber\\
{\dot H}_2 & = & - N D + 2\lambda H_2\, ,\nonumber\\
{\dot D}\  & = & -2 M H_2 + 2 N H_1\, .
\label{cevol}
\end{eqnarray}
These equations show that there are no secondary constraints, and 
that three constraints (\ref{Hamiltonian}) are first class.  The 
dynamics of the system is given entirely by the constraints and 
the Hamiltonian is $H = N H_1 + M H_2 + \lambda D$.  Since we 
have four real dynamical variables ($\vec u$ and $\vec v$) and 
three first class constraints, the system has a single physical 
degree of freedom.

The Poisson algebra of the constraints can be read directly from 
(\ref{cevol}); it is given by (cfr.\ eq.\ (\ref{hd2}))
\begin{eqnarray}
\{ H_1 \, , \, H_2 \} & = & D\, ,\nonumber\\
\{ H_1 \, , \, D \}\, & = & - 2 H_1 \, ,\nonumber\\
\{ H_2 \, , \, D \}\, & = &  2 H_2 \, .\label{Algebra}
\end{eqnarray}
This algebra is isomorphic to $sl(2,R)$, the Lie algebra of the 
group $SL(2,R)$.

{\it Analogy with GR}.  The model has a structure 
recalling GR. The analogy is transparent in the Hamiltonian 
framework, given the similar structure of the two constraint 
algebras.  In the Lagrangian framework, compare the action 
(\ref{Action}) with the Einstein-Hilbert action $S_{GR}$. Written 
in terms of the Arnowitt-Deser-Misner (ADM) variables,  $S_{GR}$ is
\begin{eqnarray}
S_{GR}[g, N, \lambda]  & = &  \int dt \int \sqrt{g}d\vec x\  N\, 
(Dg^{2}+R[g]),\nonumber\\
Dg_{ab} & = & \frac{1}{N} (\dot g_{ab} - 2D_{(a}\lambda_{b)}) 
\label{GRact}
\end{eqnarray}
where $g$ is the three-dimensional metric, $N$ the lapse and 
$\lambda$ the shift, $R$ the three-dimensional Ricci scalar, we 
have indicated the extrinsic curvature by $-Dg_{ab}$ and 
written $Dg^{2}\equiv Dg_{c}^{c}Dg_{d}^{d}-Dg_{ab}Dg^{ab}$.
Notice that the two components of $\vec u$ mimic the metric in a 
space point, the two components of $\vec v$ mimic the metric in a 
second space point, $N$ mimics the lapse in the first point, $M$ 
the lapse in a second point and $\lambda$ the shift.  The sum in  
(\ref{Action}) mimics the integration over $x$ in (\ref{GRact}), 
and the definition 
of $D\vec v$ and $D\vec v$ mimics the extrinsic curvature.  

{\it Gauge invariance}.  Under an infinitesimal gauge 
transformation generated by infinitesimal time dependent 
parameters $n(t), m(t), l(t)$, the canonical variables transforms 
as  \cite{Dirac}
\begin{eqnarray} 
\delta {\vec u} & = & l(t) {\vec u} + n(t) {\vec p}\, ,\nonumber\\
\delta {\vec p} & = & m(t) {\vec u} - l(t) {\vec p}\, ,\nonumber\\
\delta {\vec \pi} & = & l(t) {\vec \pi} + n(t) {\vec v}\, ,\nonumber\\
\delta {\vec v} & = & m(t) {\vec \pi} - l(t) {\vec v}\, , \label{Variation}
\end{eqnarray} 
while the Lagrange multipliers transform as \cite{Henneaux}
\begin{eqnarray}
\delta N & = & {\dot n}(t) - 2 n(t) \lambda + 2 l(t) N\, ,\nonumber\\
\delta M & = & {\dot m}(t) + 2 m(t) \lambda - 2 l(t) M\, ,\nonumber\\
\delta \lambda\ \, & = & \, {\dot l}(t) + n(t) M - m(t) N\, . 
\end{eqnarray}
We can check the transformation of the action (\ref{Monica}) 
under this infinitesimal variation of the canonical variables
and the Lagrange multipliers. We find that $\delta S=0$ provided that
the boundary term $n(t) (p^2 +v^2) + m(t) ({\pi}^2 +u^2)|^{t=t_f}_{t=t_i}$
vanishes. 

The problem of finding the {\it finite} gauge transformations can 
be solved by using the fact that (\ref{Variation}) is an 
infinitesimal $SL(2,R)$ transformation.  More precisely, each one 
of the four pairs $(u^{1}, p^{1})$, $(u^{2}, p^{2})$, $(\pi^{1}, 
v^{1})$, $(\pi^{2}, v^{2})$ (notice that the order is inverted in 
the second two), transforms in the fundamental representation of 
$SL(2, R)$.  It follows that the finite gauge transformation of 
the canonical variables generated by the first class constraints 
are given by finite $SL(2, R)$ transformations as follows
\begin{eqnarray}
{\vec u}'  =  \alpha(t) {\vec u} + \beta(t) {\vec p}\, ,
&\hspace{2em}&
{\vec \pi}'  =  \alpha(t) {\vec \pi} + \beta(t) {\vec v}\, , 
 \nonumber\\
{\vec p}'  = \gamma(t) {\vec u} + \delta(t) {\vec p}\, ,
&\hspace{2em}&
{\vec v}'  =  \gamma(t) {\vec \pi} + \delta(t) {\vec v}\, ,
\label{Variables}
\end{eqnarray}
where the matrix 
\begin{equation}
	G(t) = \pmatrix{\alpha(t)  & \beta(t) \cr
\gamma(t) & \delta(t) }
\label{Gt}
\end{equation}
is in $SL(2, R)$, that is, with the only restriction that 
$\alpha(t)\delta(t)-\beta(t)\gamma(t)=1\,$.  Thus, the system is 
invariant under an $SL(2, R)$ gauge invariance local in time.

The finite transformation law for the Lagrange multipliers can be 
found from the definitions of the momenta.  We obtain with some 
algebra
\begin{eqnarray}
N' & = & \alpha^2 N -\beta^2 M -2 \alpha \beta \lambda 
+ \alpha {\dot \beta} -{\dot a} \beta \, , \nonumber\\
M' & = & -\gamma^2 N + \delta ^2 M +2 \gamma \delta  \lambda 
+ {\dot \gamma} \delta  - \gamma {\dot \delta } \, , \nonumber\\
\lambda ' & = & -\alpha \gamma N + \beta \delta  M + (\alpha \delta  
+\beta \gamma)\lambda 
+ {\dot \alpha} \delta  - {\dot \beta} \gamma  \, .\label{Multipliers}
\end{eqnarray}
Below we give a clean geometric interpretation of these 
ugly-looking transformations.

We can now check the invariance of the action.  By plugging 
(\ref{Variables}) and (\ref{Multipliers}) into the action 
(\ref{Monica}) we get with some algebra 
\begin{eqnarray}
S' &=&\int dt \left [ 
{\dot{\vec u}}\cdot {\vec p} + 
{\dot {\vec v}}\cdot {\vec {\pi}}-( N H_1 + M H_2 + \lambda D) 
 \right ] + \nonumber \\
&& \hspace{2em} + \left [
(\beta\gamma) ({\vec u}\cdot {\vec p} + {\vec v} \cdot {\vec \pi} ) + 
\frac12 (\alpha\gamma) (u^2 +{\pi}^2) \right.  \nonumber \\
&& \hspace{2em} \left. + \frac12 (\beta\delta)
(p^2 +v^2) \right ]^{t=t_f}_{t=t_i}\, .
\end{eqnarray}
The action is invariant provided that the boundary term vanishes.  

{\it Solution to the equations of motion}.  The evolution of the 
system can be viewed geometrically.  Let us focus on the $(\vec 
u, \vec p)$ sector --the $(\vec \pi, \vec v)$ behaves in the same 
manner.  The equations of motion (\ref{Equations}) for this 
sector can be written in the form
\begin{eqnarray}
\frac{d}{dt}
\pmatrix{{\vec u} \cr {\vec p} \cr } - \pmatrix{
 \lambda  &  N \cr
 M & - \lambda } \pmatrix{ {\vec u} \cr {\vec p} \cr } = 0\, .
 \label{eom}
\end{eqnarray} 
The matrix composed by the Lagrange multipliers is valued in the 
Lie Algebra of the $SL(2,R)$ group and can be viewed as the 
Yang-Mills connection for the local (in time) gauge group $SL(2, 
R)$
\begin{equation}
	A(t) =  \pmatrix{
 \lambda(t)  &  N(t) \cr
 M(t) & - \lambda(t) }\, .
	\label{A}
\end{equation}
This is not a vague analogy: using this notation, the 
ugly transformation (\ref{Multipliers}) becomes  
\begin{equation}
  A' = G A G^{-1} - G \frac{d}{dt} G^{-1}
\label{conn}
\end{equation}
That is, $A$ transforms precisely as a connection.  Under a time 
dependent gauge transformation $G(t)$,\ $(\vec u, \vec p)$ 
transform as in (\ref{Variables}), $A$ transforms as in (\ref{conn}) 
and the form of the equation of motion (\ref{eom}) is preserved.

Given the geometric analogy, it is easy to integrate the 
equations of motion.  The Lagrange multipliers can be chosen as 
arbitrary functions of time, namely we can choose an arbitrary 
time dependent $sl(2,R)$ matrix $A(t)$.  The solution of the 
equations of motion (\ref{eom}) is then obtained from the initial 
value $(u_{0},p_{0})$ at time $t=0$ by
\begin{eqnarray}
\pmatrix{{\vec u}(t) \cr {\vec p}(t) \cr } =  \pmatrix{
a(t) & b(t) \cr
c(t) & d(t) } 
\pmatrix{ {\vec u}_{0} \cr {\vec p}_{0} \cr } \, , 
\end{eqnarray}
where the matrix 
\begin{equation}
	U(t) =  \pmatrix{
a(t) & b(t) \cr
c(t) & d(t) }
\end{equation}
satisfies the  parallel transport equation 
\begin{equation}
\frac{d}{dt} U(t) - A(t) U(t) = 0\, .
\label{pt}
\end{equation}
The solution is given by the time ordered exponential  
\begin{eqnarray}
U(t) =  {\cal P} e^{\int_0^{t} A(t') dt'}\, .\label{Holonomy}
\end{eqnarray}
Alternatively, we can chose $U(t)$ as an arbitrary one parameter 
(differentiable) family of $SL(2,R)$ matrices, and compute the 
Lagrange multipliers by derivation.  The dynamics of the $(\vec 
\pi, \vec v)$ sector is the same, with the same $U(t)$ (one has 
only to remember that $\vec v$ appears second in the $(\vec \pi, 
\vec v)$, unlikely $\vec u$).  This gives the complete solution 
of the classical equations of motion.

In conclusion, the general solution of the Lagrange equations is
\begin{eqnarray}
	\vec u(t) & = & a(t) \vec u_{0} + b(t) \vec p_{0}, \nonumber \\
	\vec v(t) & = & c(t) \vec \pi_{0} + d(t) \vec v_{0}. 
	\label{solution}
\end{eqnarray}
with 
\begin{equation}
	a(t)d(t)-b(t)c(t)=1
	\label{determ}
\end{equation}
and $(\vec u_{0},\vec v_{0}, \vec p_{0},\vec \pi_{0})$ satisfying 
the constraints, that is $\vec p_{0}^{2}=\vec v_{0}^{2},\ \  \vec 
\pi_{0}^{2}=\vec u_{0}^{2}$ and $\vec u_{0}\cdot\vec p_{0}=\vec 
v_{0}\cdot\vec\pi_{0}$.  The corresponding Lagrange multipliers 
are obtained from (\ref{pt}).  
\begin{eqnarray}
	N(t) & = & \dot b(t) a (t) - \dot a(t) b(t)\, ,   \\
	M(t) & = & \dot c(t) d(t) - \dot d(t) c(t)\, ,   \\
	\lambda(t) & = & \dot a(t) d(t) - \dot b(t) c(t)\, .
\end{eqnarray}
As expected for a fully constrained system, a solution of the 
equations of motion is given by a one-parameter family of gauge 
transformations.

Let us construct the general solution in a given gauge.  We 
consider the gauge $M=-1$, $N=+1$ and $\lambda=0$.  The matrix 
$A$ is then the unit antisymmetric matrix (and time independent) 
and its holonomy $U(t)$ is the rotation matrix by an angle $t$.  
We still have three arbitrary gauge fixings to impose at $t=0$.  
We choose $\vec u^{2}=\vec v^{2}$, $\vec u\cdot \vec p=0$ and 
$u^{2}(0)=0$.  Using the constraints and the general solution 
(\ref{solution}), we obtain 
\begin{eqnarray}
{\vec u}(t) & = & ( r \cos(\epsilon t) \, , \, r \sin(\epsilon t) 
)\, ,\nonumber\\
{\vec v}(t) & = & ( r \cos{(\epsilon' t +\phi)}\, , \, 
r \sin{(\epsilon' t +\phi)} )\, ,\nonumber\\
{\vec p}(t) & = & ( -r \epsilon  \sin(\epsilon  t) \, , \, r \epsilon  
\cos{(\epsilon t)} ) \, , \nonumber\\
{\vec \pi}(t) &  = &  ( r \epsilon'\sin{(\epsilon' t +\phi)}\, , \, 
- r \epsilon' \cos{(\epsilon' t +\phi)} )\, , 
\label{solutgauge}
\end{eqnarray} 
with $\epsilon=\pm 1$ and $\epsilon'=\pm 1$.  In this gauge, the 
two vectors $\vec u$ and $\vec v$ have the same length and rotate 
with the same angular speed, equal to one.  Notice that the 
solution depends on two (continuous) parameters.  $r\in R^{+}$ is 
the length of the vectors, and $\phi\in S_{1}$ is their relative 
angle at $t=0$.  Since the space of solutions is two-dimensional, 
there is a single degree of freedom, as anticipated.  In 
addition, there are the two discrete parameter $\epsilon$ and 
$\epsilon'$.  These distinguish four branches of the space of 
solutions, in which each of the two vectors rotate either 
clockwise or anti-clockwise.


\section{Observables}

{\it Dirac Observables}.  An observable is a function on the 
constraint surface that is invariant under the gauge 
transformations generated by {\em all\/} first class constraints.  
Equivalently, an observable is a function on the phase space 
which has weakly vanishing Poisson brackets with the first class 
constraints.  To find gauge invariant observables, we can proceed 
as follows.  As already noticed, (\ref{Variables}) indicates that 
the four two-dimensional vectors $\vec 
x_{i}=(x_{i}^{1},x_{i}^{2}), \ \ i=1,2,3,4$
\begin{displaymath}
\vec x_{1}= \! \pmatrix{ u^{1} \cr p^{1}}\!, \ 
\vec x_{2}= \! \pmatrix{ u^{2} \cr p^{2}}\!, \  
\vec x_{3}= \! \pmatrix{ \pi^{1} \cr v^{1}}\!, \  
\vec x_{4}= \! \pmatrix{ \pi^{2} \cr v^{2}} 
\end{displaymath}
transform under gauge transformation in the fundamental 
representation of $SL(2, R)$. But  $SL(2, R)$ preserves  
areas in $R^{2}$, that is, it preserves the vector product of any
two vectors. It follows immediately that the six observables
\begin{equation} 
O_{ij} = \vec  x_{i} \times  \vec x_{j} = 
x_{i}^{1}x_{j}^{2}-x_{i}^{2}x_{j}^{1}, 
\end{equation}
are all gauge invariant. Explicitly:
\begin{eqnarray}
O_{12} = u^{1}p^{2}-p^{1}u^{2}, &\hspace{2em}& O_{23} = 
u^{2}v^{1}-p^{2}\pi^{1}, \nonumber \\
O_{13} = u^{1}v^{1}-p^{1}\pi^{1}, &\hspace{2em}& O_{24} = 
u^{2}v^{2}-p^{2}\pi^{2}, \nonumber \\
O_{14} = u^{1}v^{2}-p^{1}\pi^{2}, &\hspace{2em}& O_{34} = 
\pi^{1}v^{2}-v^{1}\pi^{2} .  \label{O}
\end{eqnarray}
The Poisson brackets between the components of the 
$\vec x_{i}$ are
\begin{equation}
\{x_{i}^{1},x_{j}^{1}\} = 0, \ \ \ 
\{x_{i}^{2},x_{j}^{2}\} = 0, \ \ \   
\{x_{i}^{1},x_{j}^{2}\} = g_{ij}.  
\end{equation}
where $g_{ij}$ is the diagonal matrix $[1,1,-1,-1]$.  From this 
observation, it easy to compute the Poisson algebra of the 
$O_{ij}$ observables
\begin{equation} 
\{O_{ij},O_{kl}\} =  g_{ik}O_{jl} - g_{il}O_{jk}
                   + g_{jl}O_{ik} - g_{jk}O_{il} .
\label{pbo}
\end{equation}
Therefore the Poisson algebra of the six gauge invariant 
observables $O_{ij}$ is isomorphic to the Lie algebra of 
$SO(2,2)$.

Since the physical space is two-dimensional (1 degree of 
freedom), there are at most two independent continuous 
observables.  Therefore there must be four relations between the 
six observables $O_{ij}$, when the constraints are imposed.  
These relations can be easily obtained by computing the 
observables $O_{ij}$ in the gauge (\ref{solutgauge}) at $t=0$.  
In fact, a relation between gauge invariant quantities which is 
true in a particular gauge is also true in general.  From 
(\ref{solutgauge}) we have
\begin{eqnarray}
O_{12}= \epsilon J \, ,\ \ \hspace{2em} &\hspace{2em}& 
O_{34} = \epsilon' J \, , \nonumber \\
O_{13}= J \cos\phi \, ,  &\hspace{2em}&
O_{24}=   \epsilon\epsilon' J \cos\phi  \, , \nonumber \\
O_{14}=  J \sin\phi\ \, , &\hspace{2em}& 
O_{23}= - \epsilon\epsilon' J \sin\phi  \, .
\label{upsi}
\end{eqnarray} 
where we have introduced 
\begin{equation}
J= r^{2}. 
\end{equation}
Clearly 
\begin{eqnarray}
\epsilon\ O_{34}\ \ \   &=& \ \epsilon'\  O_{12}\, , \label{pm} \\
\epsilon\ O_{24}\ \ \ &=& \ \epsilon'\  O_{13}\, ,  \\
\epsilon\ O_{23}\ \ \ &=& - \epsilon'\  O_{14} \, ,   \\
O_{ij}O^{ij}&=&\ \ 0 \, , \label{Oconstr}
\end{eqnarray} 
In the last equation, indices are raised with $g_{ij}$.  Since 
the $O_{ij}$ are gauge invariant, these relations hold in general 
on the constraint surface.

Thus, the two continuous quantities $J\in R^{+}, \phi\in S_{1}$ 
and two discrete quantity $\epsilon,\epsilon'=\pm 1$, defined in 
general by (\ref{upsi}), namely by  
\begin{eqnarray}
	\epsilon & = & {u^{1}p^{2}-p^{1}u^{2}\over|u^{1}p^{2}-p^{1}u^{2}|} 
	\, , \nonumber \\
	\epsilon' & = 
	&{\pi^{1}v^{2}-v^{1}\pi^{2}\over|\pi^{1}v^{2}-v^{1}\pi^{2}|} \, , 
	\nonumber \\
	J & = & |u^{1}p^{2}-p^{1}u^{2}| \, , 
	\nonumber  \\
	\phi & = & \arctan{u^{1}v^{2}-p^{1}\pi^{2}\over
	u^{1}v^{1}-p^{1}\pi^{1}}\, , 
	\label{jpe}
\end{eqnarray}
are gauge invariant observables.  They can be taken as 
coordinates of the physical gauge-invariant phase space.  Using 
(\ref{pbo},\ref{upsi},\ref{Oconstr}), straightforward algebra 
yields the physical Poisson brackets:
\begin{equation}
	\{J ,\phi \} = \epsilon\epsilon'. 
	\label{Jpsi} 
\end{equation}
($\epsilon$ and $\epsilon'$ commute with everything.)  Notice that 
$J=0$ is a single point (whatever $\phi$, $\epsilon$ and 
$\epsilon'$).  Therefore the phase space has the topology of four 
cones connected at their vertices ($J=0$).  See Figure 1.
\begin{figure}[h]
  \centerline{\mbox{\epsfig{file=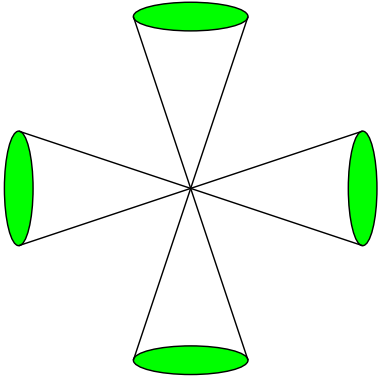}}} 
	\caption{The topology of the phase space.}
\end{figure}
Notice that 
\begin{eqnarray}
O_{12} &=& \epsilon J = \vec u \times  \vec p  \, , \nonumber \\
-O_{34} &=& -\epsilon' J = \vec v \times  \vec \pi \, .
\label{Obs}
\end{eqnarray}
are the ``angular momenta'' of the two two-dimensional 
``particles'' $\vec u$ and $\vec v$.  Since, from (\ref{pm}), 
$(O_{12})^{2}= (-O_{34})^{2}$, the two particles have the same 
``total angular momentum''.  In the gauge (\ref{solutgauge}), 
$\vec u$ and $\vec v$ rotate at equal angular speed: each one of 
the four cones represents an orientation of the two rotations, 
$J$ is their angular momentum and $\phi$ determines relative 
angle between $\vec u$ and $\vec v$.

The other four $O_{ij}$ arrange naturally in a $2\times 2$ matrix
\begin{equation}
M^{ab} \equiv \pmatrix{O_{13} & O_{14} \cr O_{23} & O_{24}} = 
u^{a}v^{b}-p^{a}\pi^{b}, 
\end{equation}
where $a,b=1,2$.  If we solve (\ref{solution}) for 
$a(t),b(t),c(t)$ and $d(t)$ and we insert the solution in 
(\ref{determ}), we obtain with some straightforward algebra
\begin{equation}
    u^a(t) v^b(t)\ \epsilon_{ac}\epsilon_{bd}\ M^{cd} =  
    O_{12}O_{34} .
\end{equation}
(The $O_{ij}$ and $M^{cd}$ observables are time independent.)  
Using (\ref{upsi}), this relation becomes
\begin{eqnarray}
 \left[ u^{1}(t)v^{1}(t)+\epsilon\epsilon' 
 u^{2}(t)v^{2}(t)\right]\cos \phi && \nonumber \\
 + \left[ u^{1}(t)v^{2}(t)-\epsilon\epsilon' 
 u^{2}(t)v^{1}(t)\right]\sin \phi &=& J.
	\label{main}
\end{eqnarray}
This is a key equation, which entirely captures the physical 
content of the model.  It expresses the relation between the 
Lagrangian variables $(\vec u, \vec v)$ in each 
$(J,\phi,\epsilon,\epsilon')$ state.  The state of the system, 
$(J,\phi,\epsilon,\epsilon')$, cannot be computed from the 
knowledge of the position $\vec u, \vec v$ at a single time: two 
times, or a time derivative, are needed, as for any dinamical 
system.  Once the state is determined, equation (\ref{main}) 
provides us with the entire gauge invariant information: the 
relation between the Lagrangian variables at any other time.

We also define the two complex conjugate observables  
\begin{eqnarray}
R := \epsilon J e^{i\phi} &=& \epsilon(O_{13}+i O_{14}) = 
\epsilon' (O_{24}-i O_{23}), \label{RS} \\
S := \epsilon J e^{-i\phi} &=& \epsilon(O_{13}-i O_{14}) = 
\epsilon' (O_{24}+i O_{23}),
\label{SR}
\end{eqnarray}
which will be convenient in the quantum theory.  A complete set 
of observables is given by $J, R, S, \epsilon, \epsilon'$ with the 
reality conditions
\begin{equation}
\overline J = J, \ \ \  \overline R =  S.
\label{reality}
\end{equation}
Clearly
\begin{equation}
\cos \phi = \frac{1}{2\epsilon}(R+S)\ J^{-1}, \ \ \  \sin \phi = 
\frac{1}{2i\epsilon}(R-S)\ J^{-1}.  
\label{cossin}
\end{equation}

{\it Evolving constants.} The physical phase space is the 
two-dimensional space of the gauge orbits on the constraint 
surface.  A point in the physical phase space is determined by 
$(J, \phi, \epsilon, \epsilon')$.  This description of the system 
resolves gauge invariance, but looses reference to time 
evolution.  Time evolution is, as in any fully constrained 
theory, a gauge transformation.

In certain fully constrained physical models such as the free 
relativistic particle or the Nambu string, there is a global 
implementation of the kinematical Poincar\'e group.  The 
generator of this group that corresponds to the energy, can be 
taken as the physical Hamiltonian for time evolution.  In other 
words, for these systems the natural time evolution can be 
introduced in the frozen reduced phase space by using the energy 
as Hamiltonian.  This provides a preferred variable that plays 
the role of time, namely of the independent evolution parameter.  
Instead, the kinematical group is absent in GR (unless additional 
structure, such as flat asymptotic infinity is added), or in the 
model studied in this paper.  In these cases, there is no 
preferred time variable.  The theory just describes --very 
democratically !-- the relative evolution of the variables, as 
functions of each other, without privileging any variable as the 
independent one.  For a detailed discussion of the physical 
meaning of this very important feature of GR, see 
\cite{Rovelli99}.

One way to express evolution in these cases, is to break gauge 
invariance.  For instance, one can impose a time dependent gauge 
fixing (the analog of $x^{0}=t$ for a relativistic particle), or 
choose a gauge at time zero and then evolve with arbitrarily 
fixed Lagrange multipliers.  This amounts to arbitrarily 
choosing one of the variables as the time variable.  

Is there, in alternative, a {\it gauge invariant} description of 
time evolution?  Are there gauge invariant observables that 
capture the dynamics of the Lagrangian variables $\vec u(t), \vec 
v(t)$?  Can we talk about a gauge-invariant dynamics, if the time 
dependence of $\vec u(t)$ and $\vec v(t)$ is a gauge 
transformation?  The answer is yes \cite{evolving}. 

In fact, the gauge invariant (or physical) content of the model 
is not the description of the evolution of the 4 real variables 
$u^{1}(t), u^{2}(t),v^{1}(t),v^{2}(t)$ in the coordinate time 
$t$, but rather the description of their evolution as functions 
of each other.  More precisely, since there are 4 variables and 
the gauge orbits are 3-dimensional, the system describes the 
motion of any {\it one} of these four variables as function of 
the other three.  In other words, once the state of the system is 
known, the dynamical model allows us to predict the value of any 
one of the four Lagrangian variables from the value of the other 
three.  This prediction is univocal and 
gauge-invariant.\footnote{The situation is exactly the same as in 
GR, where the theory does not allow us to predict the value of 
the fields at given coordinate values, or the coordinate 
positions of particles, but rather allows us to predict the value 
of general covariant quantities, such as the value of the fields 
when (and where) certain other dynamical variables have given 
values \cite{diff}.}

Each solution of the classical system, namely each point of the 
phase space determines one functional relation between the four 
variables $u^{1}(t), u^{2}(t), v^{1}(t), v^{2}(t)$.  This 
functional relation allows us to compute one of these variables 
from the value of the other three.  This functional relation is 
given by equation (\ref{main}).

The form of a gauge invariant observable describing evolution is 
therefore the following.  Let us ask what is the value $U^{1}$ of 
the observable $u^{1}$, when $u^{2}$ and $\vec v$ have assigned 
values $u^{2}=x, v^{1}=y$ and $v^{2}=z$.  In other words, we 
search an observable of the form $U^{1}=U^{1}(x,y,z; 
J,\phi,\epsilon, \epsilon')$.  Solving (\ref{main}) for $u^{1}$, and 
replacing $u^{2}, v^{1}$ and $v^{2}$ with $x, y$ and $z$, we 
obtain 
\begin{equation} 
U^{1}(x,y,z; J,\phi,\epsilon, \epsilon') = {-\epsilon' x (z\cos 
\phi- y\sin \phi) + \epsilon J\over \epsilon(y\cos \phi+ z\sin \phi)}.
\label{com}
\end{equation}
This is an ``evolving constant'' in the sense of reference 
\cite{evolving}.  For any fixed state $(J,\phi,\epsilon, 
\epsilon')$, the quantity $U^{1}(x,y,z; J,\phi,\epsilon, 
\epsilon')$, viewed as a function of $x, y$ and $z$ gives the 
evolution of $u^{1}$ as a function of the other variables.  
Viceversa, for any fixed $x,y,z$, the quantity $U^{1}(x,y,z; 
J,\phi,\epsilon, \epsilon')$, viewed as a function of 
$J,\phi,\epsilon$ and $\epsilon'$, defines a {\em gauge 
invariant\/} function on the physical phase space.  Similar 
expressions can be derived from (\ref{main}) for $u^{2}, v^{1}$ 
and $v^{2}$.
\begin{eqnarray} 
U^{2}(s,y,z; J,\phi,\epsilon, \epsilon')\!\! &=&\!\! {-\epsilon s (y\cos 
\phi+ z\sin \phi) + \epsilon J\over \epsilon'(z\cos \phi- y\sin \phi)}\, , 
\nonumber \\
V^{1}(s,x,z; J,\phi,\epsilon, \epsilon') \!\! &=&\!\!  {-z (\epsilon' x\cos 
\phi+\epsilon s\sin \phi) + \epsilon J\over \epsilon s\cos\phi- \epsilon' 
x\sin\phi}\, , \nonumber \\
V^{2}(s,x,y; J,\phi,\epsilon, \epsilon')\!\! &=&\!\!  {-y (\epsilon s\cos 
\phi- \epsilon' x\sin \phi) + \epsilon J\over \epsilon' x\cos \phi+ 
\epsilon s\sin \phi}\, ,
\label{com2}
\end{eqnarray}
where $s$ is the value of $u^{1}$.  These observables describe 
the evolution of the system {\em and\/} are gauge invariant.

{\it Time reparametrization invariance}.  The system is 
invariant under time reparametrization.  If $(\vec u(t), \vec 
v(t))$ is a solution of the equations of motion, then 
\begin{equation}
\pmatrix{\vec u'(t)\cr \vec v'(t)} = 
\pmatrix{ \vec u(f(t)) \cr \vec v(f(t))}
\end{equation} 
is also a solution.  This is immediately seen from 
(\ref{solution}) and (\ref{determ}), because 
$a(f(t))d(f(t))-b(f(t))c(f(t))=1$ follows from $a(t)d(t)-b(t)c(t)=1$.

Notice that there exist gauges in which $\vec u(t)$ evolves in 
$t$ while $\vec v(t)$ remains constant.  For instance we can 
choose $M=\lambda=0$.  In this gauge,
\begin{equation}
	A = \pmatrix{0 & N(t)\cr 0 & 0}
\end{equation}
and therefore 
\begin{equation}
	U(t) =  \pmatrix{1 & b(t)\cr 0 & 1}
\end{equation}
so that 
\begin{eqnarray}
{\vec u} & = & \vec u_{0} + b(t) \vec p_{0} ,\nonumber\\
{\vec v} & = & \vec v_{0} 
\end{eqnarray}
with $N=\dot b$.  A different example is the following.  The 
solution (\ref{solutgauge}) can be gauge transformed to the 
solution
\begin{eqnarray}
{\vec u}  =  ( u \cos{t} \, , \, u \sin{t} )\, ,
\ \ \ \ &\hspace{2em}&
{\vec v}  =  ( u\, , \, 0 )\, ,\nonumber\\
{\vec p}  =  \left (u \frac{(\cos{t} -1)}{\sin{t}} \, , \, 
u \right ) \, , 
&\hspace{2em}&
{\vec \pi}  =   ( u \sin{t}\, , \, 0)\, ,
\end{eqnarray}
where the Lagrange multipliers are $\lambda=\frac{\cos{t} -1}{\sin{t}} $, 
$M=-\frac{(\cos{t} -1)}{(\sin{t})^2} $, and $N=1$. 
Similarly, there is a gauge in which $\vec v(t)$ evolves in $t$ 
while $\vec u(t)$ remains constant.  

Notice, however, that there isn't really a ``two finger time 
reparametrization invariance'' in the system \cite{luca}, in the 
sense that it is not true that if $(\vec u(t), \vec v(t))$ is a 
solution of the equations of motion, then
\begin{equation}
\pmatrix{\vec u'(t)\cr \vec v'(t)} = 
\pmatrix{ \vec u(f_{1}(t)) \cr \vec v(f_{2}(t))}
\end{equation} 
is also a solution.  In fact, in any given time $\vec u'(t)$ and 
$\vec v'(t)$ must be connected to the same point in phase space 
by a gauge transformation, but in general it is not true that 
$a(f_{1}(t))d(f_{2}(t))-b(f_{1}(t))c(f_{2}(t))=1$ when 
$a(t)d(t)-b(t)c(t)=1$.
 

\section{Quantum Dynamics}

We work in the coordinate representation.  Elements of the 
Hilbert space are functions $\Psi({\vec u},{\vec v})$ of 
the coordinates, and the momentum operators are 
\begin{equation}
{\widehat {\vec p}}= -i\hbar\vec \nabla_{u}, \ \ \ 
{\widehat {\vec \pi}}= -i\hbar\vec \nabla_{v}.  
\label{oper}
\end{equation}
By inserting these operators in 
the constraints we obtain the Dirac quantum constraints
\begin{eqnarray}
{\widehat H}_1 & = & -\frac12 \left (  
{\hbar}^2 \Delta_{u} + \vec v^2 \right ) \, ,\nonumber\\
{\widehat H}_2 & = & -\frac12 \left (  
{\hbar}^2 \Delta_{v} + \vec u^2 \right ) \, ,\nonumber\\
{\widehat D}\ & = & -i\hbar \left ( {\vec u}\cdot \vec\nabla_{u} 
- {\vec v}\cdot \vec\nabla_{v} \right )\, .
\label{quco}
\end{eqnarray}
where $\Delta_{u}=\vec\nabla_{u}^{2} = 
{\partial^{2}\over\partial{(u^{1})^{2}}}\ 
+{\partial^{2}\over\partial{(u^{2})^{2}}}$.  In the Hamiltonian 
constraint operators ${\widehat H}_1$ and ${\widehat H}_2$ there 
is a natural ordering.  In the ``diffeomorphism'' operator 
${\widehat D}$, we have chosen the ordering that leads to the 
closure of the constraint algebra and thus the absence of 
anomalies.  We have in fact
\begin{eqnarray}
\left [ {\widehat H}_1 \, , \, {\widehat H}_2 \right ] & = &\ 
i\hbar {\widehat D}\, ,\nonumber\\
\left [ {\widehat H}_1 \, , \, {\widehat D} \right ]\ & = & 
- 2 i \hbar  {\widehat H}_1 \, ,\nonumber\\
\left [ {\widehat H}_2 \, , \, {\widehat D} \right ]\ & = & \ 
 2 i \hbar  {\widehat H}_2 \, .\label{QAlgebra}
\end{eqnarray}

The physical states, in the sense of Dirac, are in the kernel of 
all the quantum constraints.  Namely, they are defined by
\begin{eqnarray}
\left ( {\hbar}^2 \Delta_{u} + \vec v^2
\right ) \Psi({\vec u}, {\vec v}) & =& 0\, ,\nonumber\\
\left ( {\hbar}^2 \Delta_{v} + \vec u^2
\right ) \Psi({\vec u}, {\vec v}) & = & 0\, ,\nonumber\\
-i\hbar \left (
{\vec u} \cdot \vec{\nabla}_{u} - {\vec v} \cdot \vec{\nabla}_{v}
\right) \Psi({\vec u}, {\vec v}) & = & 0\, .\label{Kernel}
\end{eqnarray}
We now solve this system of coupled partial differential equations. 

We transform to polar coordinates 
\begin{equation}
{\vec u}=(u \cos{\alpha}, \ \ 
u\sin{\alpha})$, $\vec v=(v \cos{\beta}, v \sin{\beta})
\end{equation}
and we multiplying the first equation of the system by 
$u^2/\hbar^{2}$ and the second by $v^2/\hbar^{2}$.  
(\ref{Kernel}) becomes
\begin{eqnarray}
\left (
u \frac{\partial}{\partial u} u \frac{\partial}{\partial u} 
+ \frac{\partial^{2}}{\partial \alpha^{2}} +\frac{u^2 v^2}{\hbar^{2}}
\right ) \Psi(u,v,\alpha,\beta) & = & 0\, ,\nonumber\\
\left (
v \frac{\partial}{\partial v} v \frac{\partial}{\partial v} 
+ \frac{\partial^{2}}{\partial \beta^{2}} +\frac{u^2 v^2}{\hbar^{2}}
\right ) \Psi(u,v,\alpha,\beta) & = & 0\, ,\nonumber\\
\left (
u \frac{\partial}{\partial u} - v\frac{\partial}{\partial v}
\right ) \Psi (u,v,\alpha,\beta) & = & 0 \, .\label{System}
\end{eqnarray}
We search a solution by  separation of variables, by writing 
\begin{equation}
	\Psi(u,v,\alpha,\beta)= A(\alpha)\ B(\beta)\ \psi(u,v). 
\end{equation}
The first two equations in (\ref{System}) give immediately
\begin{equation}
	A(\alpha) =  e^{im_{\alpha}\alpha}, \ \ \ \ 
	B(\beta) =  e^{im_{\beta}\beta},
\end{equation}
where $m_{\alpha}$ and $m_{\beta}$ must be integer for $\Psi$ 
to be continuous.  The third equation in (\ref{System}) 
implies that
\begin{equation}
	\psi(u,v)= \psi(uv)
\end{equation}
(a function of the product $uv$).  Plugging this last result back 
into the first two equations in (\ref{System}), we find that the 
first and last terms of one equation are equal to the first and 
last terms of the second.  Therefore the two middle terms must be 
equal as well.  Therefore the two equations imply $m^2_{\alpha}= 
m^2_{\beta}$.  We put
\begin{equation}
 m_{\alpha} = \epsilon m, \ \ \ \  m_{\beta}= - \epsilon' m, 
\ \ \ \epsilon, \epsilon' =\pm 1, 
\end{equation}
with $m$ any nonnegative integer.  The minus is inserted for later 
convenience. Using this, the first two equations of the 
system become equal to each other and reduce to 
\begin{equation}
\frac{d^2 f(x)}{dx^2} + \frac{1}{x} \frac{d f(x)}{dx} +\left (
1 -\frac{m^2}{x^2} \right ) f(x)=0 \, ,
\end{equation} 
where we have written $x=uv/\hbar$ and $f(x)=\psi({\hbar}x)$.  
This is the Bessel equation.  Thus, we have solved the system 
entirely.  We conclude that the physical Hilbert space is spanned 
by the basis states $|m,\epsilon, \epsilon' \rangle$, where $m$ 
is a nonnegative integer and $\epsilon, \epsilon'=\pm 1$.  In the 
coordinate representation these states are given by
\begin{eqnarray}
\langle u,v,\alpha,\beta|m,\epsilon,\epsilon'\rangle &=&
\Psi_{m, \epsilon,\epsilon'}(u,v,\alpha,\beta) \nonumber \\
&=& 
e^{i m(\epsilon \alpha - \epsilon' \beta)} \ 
J_m\left(\frac{uv}{\hbar}\right)\, ,
\label{states}
\end{eqnarray}
where $J_{m}$ is the Bessel function of order $m$.  Notice that 
for each $m>0$ there are four states 
($\epsilon=\pm1,\epsilon'=\pm1$), but for $m=0$ there is only one 
state, since 
$|m,+,+\rangle=|m,+,-\rangle=|m,-,+\rangle=|m,-,-\rangle$ 
.\footnote{We missed this point in the first version of this 
paper.  We thank Jorma Louko for pointing out the mistake.}

{\it Quantum observables and scalar product}.  Consider the 
observables $O_{ij}$ defined in (\ref{O}).  They are gauge 
invariant, and thus have vanishing Poisson brackets with the 
constraints.  We chose the natural ordering for the corresponding 
quantum operators $\hat O_{ij}$
\begin{eqnarray}
&\hat O_{12} = \hat  u^{1}\hat p^{2}-\hat p^{1}\hat u^{2}, 
\hspace{2em}& 
\hat O_{23} = \hat  u^{2}\hat v^{1}-\hat p^{2}\hat \pi^{1}, \nonumber \\
&\hat O_{13} = \hat  u^{1}\hat v^{1}-\hat p^{1}\hat \pi^{1}, 
\hspace{2em}& 
\hat O_{24} = \hat  u^{2}\hat v^{2}-\hat p^{2}\hat \pi^{2}, \nonumber \\
&\hat O_{14} = \hat  u^{1}\hat v^{2}-\hat p^{1}\pi^{2}, 
\hspace{2em}& 
\hat O_{34} = \hat \pi^{1}\hat v^{2}-\hat v^{1}\hat \pi^{2} .  
\label{Oq}
\end{eqnarray}
It is easy to see that the commutators of these operators with 
the quantum constraints (\ref{quco}) vanish.  Therefore these 
operators are well defined on the space of the solutions of the 
quantum constraints, namely on the states (\ref{states}).  We  
compute their action on these states.  Going to polar coordinates 
we see immediately that
\begin{eqnarray}
	\hat O_{12}\ \Psi_{m,\epsilon,\epsilon'} &=& 
	-i\hbar\frac{\partial}{\partial\alpha} 
	\Psi_{m,\epsilon,\epsilon'} = \epsilon m \hbar\ 
	\Psi_{m,\epsilon,\epsilon'} , \nonumber \\
	\hat O_{34}\ \Psi_{m,\epsilon,\epsilon'} 
	&=&\ \  i\hbar\frac{\partial}{\partial\beta}
	\Psi_{m,\epsilon,\epsilon'} =
	\epsilon' m\hbar \ \Psi_{m,\epsilon,\epsilon'}.  
\end{eqnarray}
Thus in the physical state space we have $\epsilon' \hat O_{12} = 
\epsilon \hat O_{34}$: the relation between the two is precisely 
the same as in the classical theory, eq.\,(\ref{pm}).  We can 
thus identify the $\epsilon$ and $\epsilon'$ appeared in the 
quantum theory with the $\epsilon$ and $\epsilon'$ appeared in 
solving the classical theory, and we conclude, from equation 
(\ref{upsi}), that the quantum operator corresponding to the 
gauge invariant observable $J$ is
\begin{equation} 
    \hat J \ |m, \epsilon,\epsilon'\rangle = 
    \hbar m\ |m, \epsilon,\epsilon'\rangle.  
\end{equation} 
Thus in the quantum theory $J$ is discrete, quantized in 
multiples of $\hbar$
\begin{equation}
	J = m\ \hbar. 
\end{equation}

Using the Bessel equation and the properties
\begin{eqnarray}
 J_{m-1}(x) &=& \frac{m}{x}J_m(x) + 
 \frac{d}{dx}J_{m}(x), \nonumber \\
 J_{m+1}(x) &=& \frac{m}{x}J_m(x) - 
 \frac{d}{dx}J_{m}(x)
\end{eqnarray}
of the Bessel functions, a straightforward but long 
calculation yields 
\begin{eqnarray}
(\hat O_{13}+i\hat O_{14})\ \Psi_{m,\epsilon,\epsilon'} &=& \epsilon 
\hbar m\  
\Psi_{m+\epsilon\epsilon',\epsilon,\epsilon'},\nonumber \\
(\hat O_{24}-i\hat O_{23})\ \Psi_{m,\epsilon,\epsilon'} &=& 
\epsilon' \hbar  m\ \Psi_{m+\epsilon\epsilon',\epsilon,\epsilon'}.
\end{eqnarray}
Thus, the quantum operator corresponding to the observable $R$ 
defined in (\ref{RS}) is 
\begin{equation}
\hat R \ |m,\epsilon,\epsilon'\rangle = \hbar m\ |m+\epsilon\epsilon',
\epsilon,\epsilon'\rangle. 
\end{equation}
In the same manner, from (\ref{SR}) we obtain 
\begin{equation}
\hat S \ |m,\epsilon,\epsilon'\rangle = \hbar m\ |m-\epsilon\epsilon', 
\epsilon,\epsilon'\rangle. 
\end{equation}
To complete the construction of the Hilbert space of the physical 
quantum states, we have to determine the scalar product on the 
space spanned by the states $|m,\epsilon,\epsilon'\rangle$.  This 
is determined by the requirement that real classical observables 
be self adjoint.  The observables $J, \epsilon$ and $\epsilon'$ 
are real, and thus we require $\hat J, \hat\epsilon$ and $\hat 
\epsilon'$ to be self adjoint.  It follows that the states 
$|m,\epsilon,\epsilon'\rangle$ which are their eigenstates must 
form an orthogonal basis.  This fixes the scalar product up to 
the norm of the basis states.  Define
\begin{equation} 
\langle m, \epsilon,\epsilon' |m,\epsilon,\epsilon'\rangle = 
c_{m,\epsilon,\epsilon'}.
\end{equation} 
Next, $S$ is the complex conjugate of $R$. Thus we require 
that $R^{\dagger}=\hat S$. It follows 
\begin{equation}
\langle m, \epsilon,\epsilon'|\ R^{\dagger}\ |n, \epsilon,\epsilon' 
\rangle = \langle m, \epsilon,\epsilon' |\ \hat S\ |n, 
\epsilon,\epsilon' \rangle.
\end{equation}
{}From which, we have easily 
\begin{equation}
  c_{m,\epsilon,\epsilon'}= c m. 
\end{equation}
Here $c$ is a positive overall normalization constant that has no 
effect on the physics, and we chose equal to 1.  This fixes the 
normalization of the orthogonal basis states, and therefore 
determines the scalar product completely.  Notice that the state 
$|0,\epsilon, \epsilon'\rangle$ has zero norm.  (This was first 
realized by Jorma Louko).  We can therefore discard it, because 
its presence has no physical consequences.  More precisely, we 
identify the $m=0$ state with the state zero.

The peculiar behavior of the $m=0$ sector of the quantum theory 
reflects the pathological properties of the corresponding 
classical state.  The quantum state $m=0$ has vanishing angular 
momentum $J$; the classical state with vanishing angular momentum 
is the (common) vertex of the four cones that form the reduced 
phase space (see Figure 1).  This is a point at which the reduced 
phase space fails to be a manifold.  Physically, this corresponds 
to the fact that small perturbations of the $J=0$ solution form 
disjoint spaces.

Thus, the quantum theory is completely defined by the states
\begin{equation} 
|\psi\rangle = \sum_{m=1,\infty; \epsilon,\epsilon'=\pm}
\ c_{m,\epsilon,\epsilon'}\ |m,\epsilon,\epsilon'\rangle,  
\end{equation}
the scalar product
\begin{equation}
\langle m,\epsilon,\epsilon' | \tilde m, \tilde \epsilon,
 \tilde \epsilon' \rangle 
= m \ \delta_{m, \tilde m}\delta_{\epsilon, \tilde \epsilon}
\delta_{\epsilon', \tilde \epsilon'}, 
\end{equation}
and the operators 
\begin{eqnarray} 
\hat J\ |m,\epsilon,\epsilon'\rangle &=& \hbar m\ 
|m,\epsilon,\epsilon'\rangle, \nonumber \\
\hat R\ |m,\epsilon,\epsilon'\rangle &=& \hbar m\ |m+\epsilon\epsilon' 
,\epsilon,\epsilon'\rangle, \nonumber \\
\hat S\ |m,\epsilon,\epsilon'\rangle &=& \hbar m\ |m-\epsilon\epsilon' 
,\epsilon,\epsilon'\rangle, \nonumber \\
\hat \epsilon\ |m,\epsilon,\epsilon'\rangle &=& \epsilon\ \ 
|m,\epsilon,\epsilon'\rangle, \nonumber \\
\hat \epsilon'\ |m,\epsilon,\epsilon'\rangle &=& \epsilon'\ 
|m,\epsilon,\epsilon'\rangle. 
\end{eqnarray} 
where it is understood that $|0,\epsilon,\epsilon'\rangle=0$. 
(That is, for instance, $\hat R\ |1,+,-\rangle= 0$).

{\it Quantum evolving constants}. 
In order to quantize the evolving constant of motion 
(\ref{com},\ref{com2}), we must construct the operators 
corresponding to the classical observables $\cos \phi$ and 
$\sin\phi$.  We denote these operators $\widehat{\cos\phi}$ and 
$\widehat{\sin\phi}$, with a slight abuse in notation.  (The 
operator $\hat \phi$ is ill defined because $\phi$ is an angle 
--see for instance \cite{ref}-- and we must deal with periodic 
functions of $\phi$ in order to have continuity all around the 
circle.)  Choosing the natural ordering given in (\ref{cossin}), 
we have immediately 
\begin{eqnarray}
\widehat{\cos\phi} \ |m,\epsilon,\epsilon'\rangle &=& \frac{ 1 
}{2\epsilon }( |m+\epsilon\epsilon', \epsilon,\epsilon'\rangle+ 
|m-\epsilon\epsilon', \epsilon,\epsilon'\rangle) ,\nonumber \\
\widehat{\sin\phi} \ |m,\epsilon,\epsilon'\rangle &=& \frac{ 1 
}{2\epsilon i}( |m+\epsilon\epsilon', \epsilon,\epsilon'\rangle - 
|m-\epsilon\epsilon', \epsilon,\epsilon'\rangle)
\end{eqnarray}
(where, again, it is understood that 
$|0,\epsilon,\epsilon'\rangle=0$.) 

A convenient representation of the theory can be 
obtained by representing a generic state 
\begin{equation}
|\psi\rangle = \sum_{m, \epsilon, \epsilon'} 
\psi_{m,\epsilon,\epsilon'}\ |m,\epsilon,\epsilon'\rangle 
\end{equation}
by the four functions on $S_{1}$ 
\begin{equation}
\psi_{\epsilon,\epsilon'}(\phi) = \sum_{m=1}^{\infty} 
\psi_{m,\epsilon,\epsilon'} \ 
e^{i\epsilon\epsilon'm(\phi+\frac{\pi}{2}(3+\epsilon))}. 
\label{sum}
\end{equation} 
The scalar product turns out to be 
\begin{equation} 
\langle \psi_{\epsilon,\epsilon'} | 
\tilde\psi_{\epsilon,\epsilon'} \rangle = -i 
\epsilon\epsilon'\int d\phi \ \bar 
\psi_{\epsilon,\epsilon'}(\phi)\ \frac{d}{d\phi} \ 
\tilde\psi_{\epsilon,\epsilon'}(\phi).
\label{scpr}
\end{equation}
Notice that since the sum in (\ref{sum}) is restricted to $m>0$, 
the Hilbert space is formed by ``right moving'' functions 
$\psi_{+,+}(\phi)$ and $\psi_{-,-}(\phi)$, and ``left moving'' 
functions $\psi_{+,-}(\phi)$ and $\psi_{-,+}(\phi)$ only.  On 
these functions, the scalar product (\ref{scpr}) is positive 
definite.  In particular, the zero modes 
$\psi_{\epsilon,\epsilon'}(\phi)=constant$ do not belong to the 
Hilbert space.  We denote the projector that projects out the 
zero modes as $P$.  The observables are then
\begin{eqnarray} 
\hat J\ \ \psi_{\epsilon,\epsilon'}(\phi) &=& -i\hbar\epsilon\epsilon'
\, \frac{d}{d\phi}\psi_{\epsilon,\epsilon'}(\phi) , \nonumber \\
\widehat{\cos\phi}\ \psi_{\epsilon,\epsilon'}(\phi)  &=& P \ \cos\phi\ 
\psi_{\epsilon,\epsilon'}(\phi) , \nonumber \\ 
\widehat{\sin\phi}\ \psi_{\epsilon,\epsilon'}(\phi)  &=& P \ \sin\phi\ 
\psi_{\epsilon,\epsilon'}(\phi).  
\end{eqnarray} 
In this representation it is easy to write the quantum operator 
corresponding to the evolving constant of motion, which quantizes 
the observable (\ref{com}).  This is given by
\begin{eqnarray} 
&& \hat U^{1}(x,y,z) = \nonumber \\
&& P{-\epsilon\epsilon' \over y\cos \phi+ z\sin \phi} P
\left[x(z\cos \phi - y\sin \phi) + i\hbar \frac{d}{d\phi} \right] 
\nonumber
\end{eqnarray}
where we have arbitrarily picked an ordering.\footnote{General 
procedures for systematically ordering observables exist 
\cite{ordering}, and should presumably be used here.} The 
expectation value of this operator on a state 
$\Psi_{\epsilon,\epsilon'}(\phi)$ --taken with the scalar product 
(\ref{scpr})-- gives the physical mean value of the variable 
$u^{1}$ at the moment in which the three variables $u^{2},v^{1}$ 
and $v^{2}$ have value $x,y$ and $z$ (see \cite{evolving}).  
Similar operators can be defined for the three other evolving 
constants (\ref{com2}).


\section*{Acknowledgments} We are very indebited to Jorma Louko 
for pointing out a problem in the first version of the paper 
(regarding the $m=0$ state) and for the discussion of the 
physical scalar product.  We thank Sameer Gupta, Laurent Freidel, 
John Baker, Raymond Puzio and the other postdocs and students at 
the Center for Gravitational Physics at Penn State University for 
comments and help, and in particular for first realizing that the 
symmetry of the model is local $SL(2,R)$; Roberto De Pietri for 
pointing out references \cite{luca,ordering}.  MM's postdoctoral 
fellowship at the Department of Physics and Astronomy of the 
University of Pittsburgh is funded through the CONACyT of Mexico, 
fellow number 91825.  Also MM thanks financial support provided 
by the {\it Sistema Nacional de Investigadores} of the 
Secretar\'{\i}a de Educaci\'on P\'ublica (Mexico).  This work has 
been supported also by NSF Grant PHY-95-15506.


\end{document}